\documentclass[]{aastex63}
\usepackage{amsmath}

\submitjournal{ApJ}

\shorttitle{Magnetic switchback formation}
\shortauthors{Magyar et al.}

\begin{document}

\title{Could switchbacks originate in the lower solar atmosphere? \\
I. Formation mechanisms of switchbacks}

\correspondingauthor{Norbert Magyar}
\email{norbert.magyar@warwick.ac.uk}

\author[0000-0001-5731-8173]{Norbert Magyar}
\affiliation{Centre for Fusion, Space and Astrophysics, \\Physics Department, University of Warwick,\\ Coventry CV4 7AL, UK}
\affiliation{Centre for mathematical Plasma Astrophysics (CmPA),\\
KU Leuven, \\
Celestijnenlaan 200B bus 2400, B-3001 Leuven, Belgium}

\author[0000-0002-0061-5916]{Dominik Utz}
\affiliation{Computational Neurosciences, Neuromed Campus, Kepler University Hospital,\\ 4020 Linz, Austria}
\affiliation{Instituto de Astrofísica de Andalucía IAA-CSIC,\\
Glorieta de la Astronomía s/n,\\
18008 Granada, Spain}


\author[0000-0003-3439-4127]{Robertus Erd\'elyi}
\affiliation{Solar Physics \& Space Plasma Research Center (SP2RC),\\
School of Mathematics and Statistics, University of Sheffield, \\
Hicks Building, Hounsfield Road, S3 7RH, UK}
\affiliation{Dep. of Astronomy, E\"otv\"os Lor\'and Univ.,\\
P\'azm\'any P. s\'et\'any 1/A,\\
Budapest, H-1117, Hungary}
\affiliation{Gyula Bay Zolt\'an Solar Observatory (GSO),\\
Hungarian Solar Physics Foundation (HSPF), \\
Pet\H{o}fi t\'er 3., Gyula, H-5700, Hungary}

\author[0000-0001-6423-8286]{Valery M. Nakariakov}
\affiliation{Centre for Fusion, Space and Astrophysics, \\Physics Department, University of Warwick,\\ Coventry CV4 7AL, UK}
\affiliation{School of Space Research, Kyung Hee University, Yongin, 17104, \\Republic of Korea}




\begin{abstract}
The recent rediscovery of magnetic field switchbacks or deflections embedded in the solar wind flow by the Parker Solar Probe mission lead to a huge interest in the modelling of the formation mechanisms and origin of these switchbacks. Several scenarios for their generation were put forth, ranging from lower solar atmospheric origins by reconnection, to being a manifestation of turbulence in the solar wind, and so on. 
Here we study some potential formation mechanisms of magnetic switchbacks in the lower solar atmosphere, using three-dimensional magneto-hydrodynamic (MHD) numerical simulations. The model is that of an intense flux tube in an open magnetic field region, aiming to represent a magnetic bright point opening up to an open coronal magnetic field structure, e.g. a coronal hole. The model is driven with different plasma flows in the photosphere, such as a fast up-shooting jet, as well as shearing flows generated by vortex motions or torsional oscillations.
In all scenarios considered, we witness the formation of magnetic switchbacks in regions corresponding to chromospheric heights. Therefore, photospheric plasma flows around the foot-points of intense flux tubes appear to be suitable drivers for the formation of magnetic switchbacks in the lower solar atmosphere. Nevertheless, these switchbacks do not appear to be able to enter the coronal heights of the simulation in the present model. In conclusion, based on the presented simulations, switchbacks measured in the solar wind are unlikely to originate from photospheric or chromospheric dynamics.   

\end{abstract}

\keywords{solar magnetic fields, MHD simulations, switchbacks, Parker Solar Probe}

\section{Introduction}\label{sec:intro}
With the advent of the Parker Solar Probe (PSP for short) we are able to directly measure properties of the solar wind closer to the Sun than ever before \citep{2016SSRv..204....7F}. One of the early highlight discoveries of PSP is the omnipresence of strong local deflections of the magnetic field in the solar wind, mostly referred to as switchbacks \citep{2019Natur.576..228K,2020ApJS..246...45H}. These structures, also called as folds, jets, or spikes, are not necessarily full reversals of the local magnetic field, unlike some names suggest. In fact, the measured distribution of deflections resembles a power law, with most of the deflections at small angles relative to the Parker spiral \citep{2020ApJS..246...39D}. There is evidence that switchbacks are localized kinks in the magnetic field and not polarity reversals or closed loops \citep{2020ApJS..246...74W,2020ApJS..246...67M}. However, switchbacks are not a new finding. They were already observed for several decades, by e.g. \textit{Helios} \citep[e.g.,][]{2018MNRAS.478.1980H}, out to 1.3 AU by \textit{Ulysses} \citep[e.g.,][]{1999GeoRL..26..631B,2013AIPC.1539...46N}. The novelty in the PSP observations is their sharpness and omnipresence \citep[among other features, see][]{2020ApJS..246...39D}, indicating that switchbacks are a more frequent feature closer to the Sun. This observation lends the possibility that switchbacks may originate lower in the solar atmosphere. The formation mechanism(s) of switchbacks, and whether they represent large-amplitude Alfv\'en waves or structures advected by the solar wind, is as of yet not known. Several explanations about their origin have been put forth. Among these theories, interchange reconnection is the most focused upon \citep{2004JGRA..109.3104Y,2005ApJ...626..563F,2020ApJ...894L...4F}. In this scenario, switchbacks are generated in the solar corona, and form at the reconnection sites between open and closed magnetic fluxes. In some studies, it is argued that the interchange reconnection results in magnetic flux ropes, which are ejected by the reconnection outflow, and are advected by the wind  \citep{2020arXiv200905645D}. In other studies, reconnection is thought to generate either Alfv\'enic \citep{2020arXiv200909254H} or fast magnetosonic \citep{2020ApJ...903....1Z} wave pulses. Alternatively, as the distribution of deflections appears to be featureless and monotone in switchbacks \citep{2020ApJS..246...39D}, they may not be a distinct feature but a manifestation of the ensuing turbulent dynamics in the solar wind \citep{2020ApJ...891L...2S}. Thus, it is still not clear whether switchbacks originate in the lower solar atmosphere, or represent dynamic features of solar wind turbulence. Additionally, it is not clear whether they are wavelike perturbations, propagating at the Alfv\'en or some other characteristic speed, or structures that are advected by the wind, such as flux ropes. However, some observations offer good constraints on the nature of switchbacks. Given that  switchbacks are characterized by a strong Alfv\'enic correlation of their velocity and magnetic field perturbations, a nearly constant magnetic pressure, and velocity enhancements along the propagation direction, an interpretation in terms of propagating nonlinear Alfv\'en waves is a plausible scenario \citep{2014GeoRL..41..259M}. Alternatively, their localization in the perpendicular direction and the kink-like geometry may indicate their kink fast magnetoacoustic nature \citep[e.g.,][]{2008ApJ...676L..73V}. An additional option is the association of the switchbacks with kink solitons which keep their shape because of the balance between nonlinear and dispersive effects \citep[see, e.g.,][]{2010PhPl...17h2108R}, i.e. a stationary nonlinear kink wave pulse. Other observations, such as a sharp rise in ion temperature at the boundaries of switchbacks are more compatible with an origin by reconnection \citep{2020ApJS..246...68M,2020ApJS..249...28F}. \par 
Judging by these observed constraints, a lower solar atmospheric origin and a nonlinear Alfv\'enic nature of the switchbacks seem likely. In this sense, any potential theory or model aimed to  elucidate their observation out to $\approx 35\ \mathrm{R}_\odot$ and beyond, should explain their generation mechanism as well as their ability to travel several solar radii without disentangling. \citet{2006GeoRL..3314101L} showed through numerical simulations that a 2.5D switchback embedded and propagating in a uniform corona is prone to unfolding, and therefore unlikely to survive out to the observed distances. Other simulations of interchange reconnection-created switchbacks show that while strongly twisted nonlinear Alfv\'en waves form as part of the process, these untwist rapidly as they propagate away from the reconnection site \citep{2018ApJ...852...98W}. The model of  \citet{2006GeoRL..3314101L} was re-visited recently by \citet{2020ApJS..246...32T}. They pointed out that if the embedded switchbacks are instead of constant magnetic pressure, as it is observed to be the case, and unlike in the  \citet{2006GeoRL..3314101L} simulations, they might indeed originate in the lower solar corona and could survive out into the solar wind. However, this conclusion also presumes a background solar wind without significant gradients in density, flow speed, etc. \par 
To date, as presented above, most of the modeling on switchback origins focused on interchange reconnection manifesting in the solar corona. However, there are presumably many other processes in the lower solar atmosphere that might generate folded magnetic fields that would propagate outward as switchbacks. In the small plasma-beta lower corona the magnetic field lines are hardly bendable by non reconnection-related processes, such as waves, flows or instabilities. The measured transverse wave amplitudes in the lower solar corona, by direct imaging \citep[e.g.,][]{2015NatCo...6E7813M} and Doppler shifts \citep[e.g.,][]{2007Sci...317.1192T}, or inferred by the amount of non-thermal broadening \citep[e.g.,][]{2012ApJ...751..110B}, are 30-50 times smaller than the local Alfv\'en speed. Instabilities, such as e.g. the Kelvin-Helmholtz instability, require super-Alfv\'enic shear flows along the field \citep{1961hhs..book.....C,2015ApJ...813..123Z}. While it is true that transverse waves propagating outward grow in amplitude \citep[e.g.,][]{2001A&A...374L...9M} and could reach Alfv\'enic Mach amplitude values, switchbacks generated in this way would enter the in-situ turbulence-generated scenario, and not the lower atmospheric one. Turbulence is necessary in this case as Alfv\'en waves are stable to the Kelvin-Helmholtz instability \citep{2015ApJ...813..123Z}. Note that the fastest coronal jets observed, which are reconnection-related, were able to reach Alfv\'enic Mach speeds, however their median speed is significantly lower \citep{2016SSRv..201....1R}. Nevertheless, there is an interesting possibility that at least some coronal jets could induce switchbacks. In the lower solar atmospheric layers, as the plasma beta rises to unity or above, the flows and waves induced by the convective hydrodynamic buffeting \citep[e.g.,][]{2011ApJ...730L..24K} below might readily generate switchbacks. Strong shear flows and upflows \citep[][]{2008ApJ...687L.131B,2019A&A...632A..97L,2019NatCo..10.3504L}, at a significant portion of the Alfv\'en speed, were previously observed. Particularly, flows in and around photospheric magnetic  bright points \citep[MBPs;][]{2001ApJ...560.1010B,2012ApJ...752...48C,2014ApJ...796...79U}, which then expand into an open coronal magnetic field are relevant, as these represent direct magnetic connections between the photosphere, the corona, and out to the solar wind \citep[see, eg.,][]{2019A&A...629A..22H}.
MBPs are kG strong magnetic flux concentrations in the photosphere representing the cross-section of flux tube elements \citep[][]{2013A&A...554A..65U}. They feature very short lifetimes in the range of a few minutes and are highly dynamic \citep[][]{2018ApJ...856...17L}, continuously changing their morphology under the pushing and buffeting by the surrounding plasma flows. MBPs are excellent channels for the creation of MHD waves \citep[e.g.,][]{2009Sci...323.1582J,2013A&A...559A..88S} and therefore a likely spot in searching for photospheric creation mechanisms of switchbacks. During the convective collapse process \citep[see, e.g.,][]{1979SoPh...61..363S,2008ApJ...677L.145N} strong downflows are observed within the magnetic flux element leading in a first step to an evacuation of the plasma and intensification of the magnetic field. This process is often followed by strong upflows developing into shock fronts which ultimately can disperse the magnetic field and destroy the MBP \citep[][]{2001ApJ...560.1010B}. These flows might induce various instabilities and ultimately magnetic field switchbacks. Such flows have been previously investigated in numerical simulations by \citet[e.g.,][]{2011ApJ...730L..24K}, who found that the massaging of the flux tube elements by downstreaming granular material next to the flux tube surface leads to jet-like outflow phenomena and upflows within the flux element. Moreover, there is observational evidence of the existence of small-scale vortex and shear flows associated with MBPs \citep[][]{2012ASPC..463..107S,2019NatCo..10.3504L}. Thus, MBPs appear to be a suitable photospheric feature in which the generation of switchbacks could occur. Furthermore, in case these structures are situated below coronal holes, there is a direct magnetic connectivity between them and the higher solar atmosphere and ultimately the solar wind. \par 
The generation of switchbacks is only one part of the problem. Whether these switchbacks can escape the lower atmospheric layers and are able to propagate and survive out into the solar wind is the second and arguably the less known and less understood part of the problem. In this paper, we explore through 3D MHD numerical simulations whether flows at the base of a model MBP, both transverse and along the field, can create switchbacks. Furthermore, we investigate whether these switchbacks can escape the lower layers of the solar atmosphere and become propagating structures in the corona. The paper is structured as follows. In Section~\ref{sec:model}, we present the numerical model and the setup, in Section~\ref{sec:results} the results are presented, and in Section~\ref{sec:concl} some discussion on the results and conclusions are drawn. 

\section{Numerical model}\label{sec:model}
\begin{figure}[h]
    \centering     
        \includegraphics[width=1.0\textwidth]{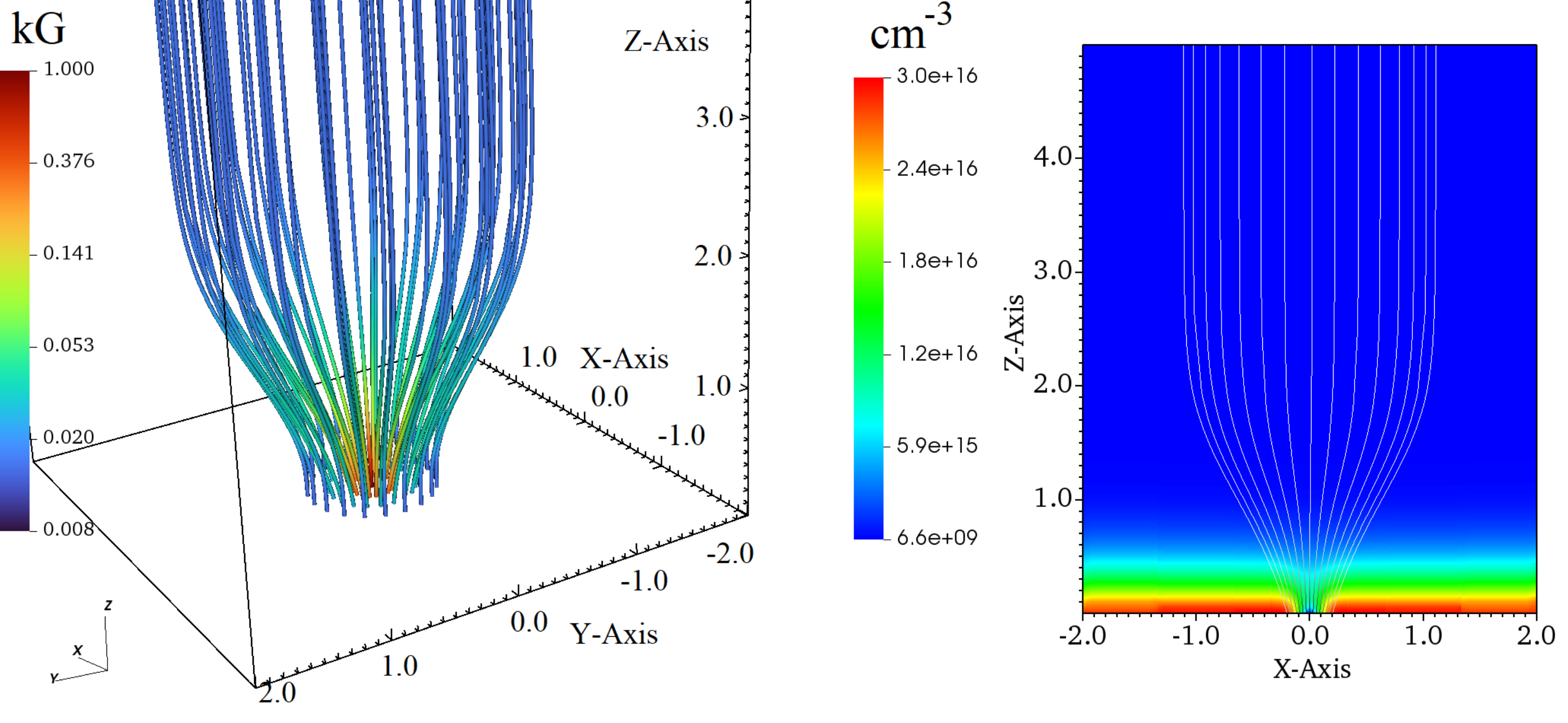}  
        \caption{Snapshots of the initial condition in the simulation. \textit{Left:} Three-dimensional stream plot of the magnetic field lines. The coloring of the field lines represents magnetic field intensity, in kG. \textit{Right:} Plot of density in a slice along the y-axis, through the center of the flux tube. The color bar is in units of number density ($\mathrm{cm^{-3}}$). Magnetic field lines are also plotted. Axis values are in $\mathrm{Mm}$.}
        \label{initcond}
\end{figure}

We run full 3D, ideal MHD numerical simulations using \texttt{MPI-AMRVAC} \citep{2014ApJS..214....4P,2018ApJS..234...30X}. A finite-volume three-step \texttt{hll} solver is employed, with \texttt{woodward} slope limiter. The solenoidality of the magnetic field is maintained by using a constrained transport method. The initial condition consists of an intense vertical flux tube embedded in the stratified lower solar atmosphere (see Fig.~\ref{initcond}).
The flux tube is prescribed using an axisymmetric similarity solution \citep[see, e.g.,][]{1958IAUS....6..263S,1965ApJ...141..548D}, with the magnetic field components given by:
\begin{equation}
  \begin{array}{lr}
      B_r(r,z,t=0) = \frac{1}{2} B_0 r e^{r^2 \left(-e^{-z}\right)-z}, \\ 
      B_z(r,z,t=0) = B_0 e^{r^2 \left(-e^{-z}\right)-z} + B_{z0},
  \end{array}
  \label{magfield}
\end{equation}
where $B_0 = 1\ \mathrm{kG}$ is the flux density at the foot-point of the flux tube in the photosphere, and $B_{z0} = 15\ \mathrm{G}$ is a background vertical magnetic field to which the flux tube expands to in the corona. We set the width of the flux tube in the photosphere to 300 km. This is around the average value of observed MBP widths \citep[e.g.,][]{2009A&A...498..289U,2018ApJ...856...17L}. The novelty in our approach in prescribing a magneto- and hydrostatic equilibrium to a similarity solution, compared to other studies in which numerical integration is employed \citep[e.g.,][]{2005A&A...441..337S,2011ApJ...727...17F}, is to find analytical solutions for the density and pressure corrections to the background atmosphere:
\begin{equation}
    \begin{array}{lr}
          p(r,z,t=0) = p_0(z) + \frac{1}{16} B_0^2 e^{-2 \left(r^2 e^{-z}+z\right)} \left(-2r^2+e^{z}-8\right), \\
         \rho(r,z,t=0) = \rho_0(z) -\frac{B_0^2 e^{-2 \left(r^2 e^{-z}+z\right)} \left(2r^2-e^{z}+16\right)}{16 g},
    \end{array}
\end{equation}
where $g = 274\ \mathrm{m\ s^{-2}}$ is the gravitational acceleration at the surface of the Sun, and $p_0(z), \rho_0(z)$ is the background hydrostatic atmosphere prescribed using the temperature distribution $T(z)$ from \citet{2007ApJ...667.1243F}, by numerically integrating the hydrostatic equation:
\begin{equation}
      \frac{d p_0(z)}{dz} = -g\rho_0(z),\ \rho_0(z) = \frac{p_0(z)m}{k_B T(z)},\\
      \label{hydrostat}
\end{equation}
where $m \approx 0.62\ m_p$ is the mean mass per particle for photospheric abundances, $m_p$ is the proton mass, and $k_B$ is Boltzmann's constant. In Eq.~\ref{hydrostat}, we assumed an ideal equation of state. 
Note that the pressure and density corrections result in an evacuation of the interior of the flux tube, as compared to the hydrostatic equilibrium outside of it, visible in Fig.~\ref{initcond}. The evacuation is height-dependent. At $100\ \mathrm{km}$ from the base of the simulation, it is a factor of $\approx 2.5$ in density and $\approx 1.6$ in pressure, respectively. Therefore the interior of the flux tube is slightly hotter than the surrounding plasma. The plasma-beta is around unity at the center of the flux tube at the same height, while outside the flux tube it is $\approx 6000$. Along the axis of the flux tube, plasma-beta reaches a maximum of $\approx 6$ at $1.5\ \mathrm{Mm}$, after which it decreases to unity again at $2.5\ \mathrm{Mm}$. At a height of $\approx 3.5\ \mathrm{Mm}$, just inside the corona, the plasma-beta reduces to $\approx 0.3$. For the implementation in the numerical simulation, the equations above are transformed from cylindrical to Cartesian coordinates, as the simulations use a Cartesian grid. The $(x,y,z)$ extents of the 3D numerical box are $(-2,2)^2\times(0,10)\ \mathrm{Mm}$, with $z=0\ \mathrm{Mm}$ being the photospheric height. The numerical grid initially consists of $48^2\times96$ numerical cells, with 5 levels of refinement. Thus the effective resolution is $768^2\times1536$, or $\approx 5.2^2 \times 6.5\ \mathrm{km}$. Refinement is enforced inside the lower flux tube (for $z \leq 750\ \mathrm{km}$ and within a radius $r \leq 750\ \mathrm{km}$ from $(x,y) = (0,0)$ at all times. Outside of this region, the refinement criteria is the presence of negative $B_z$ values, the vertical component of the magnetic field, indicating switchback forming regions. The boundary conditions are the following. Laterally, all variables obey a zero-gradient, continuous condition, which allows perturbation to freely leave the domain. In the top and bottom $z$-axis boundaries, the density and pressure are extrapolated by using Eq.~\ref{hydrostat}, with the temperature in the cell closest to the boundary. The magnetic field is extrapolated in all boundaries with the zero normal gradient and zero divergence conditions. In the top $z$-axis boundary, all velocity components are obeying the zero gradient condition. In the bottom $z$-axis boundary, velocity components that are not driven are set to anti-symmetric with respect to the boundary. We employ two different types of velocity drivers at the bottom $z$-axis boundary. First, we consider perturbations along the magnetic field, in the middle of the flux tube, of the following form: 
\begin{equation}
    v_z(r,t) = 
    \begin{cases} 
      A\ e^{-(t/\tau)^4}e^{-(r/r_b)^6} &r \leq r_b\ \mathrm{km}, \\
      0 &r > r_b\ \mathrm{km}, \\
    \end{cases}
    \label{vzdrive}
\end{equation}
where $A$ is the perturbation amplitude, $\tau$ is the characteristic driving time, and $r_b$ is the width of the perturbation in the horizontal plane. Second, we consider torsional flows, in the horizontal plane of the flux tube:
\begin{equation}
    v_\theta(r,t) = f(t) \cdot A\ e^{-\left((r-r_b)/r_b\right)^2}
    \label{vthdrive}
\end{equation}
where $r_b$ now represents the radius at which the flow amplitude is maximal, and for $f(t)$ we choose either a sinusoidal oscillating flow ($f(t) = \mathrm{sin}(2\pi t/\tau)$), or a decaying vortex flow ($f(t) = e^{-(t/\tau)^4}$), with $\tau$ either the oscillation period or the characteristic driving time. The width of the Gaussian is chosen in such a way as to drive $v_\theta$ only a narrow ring around $r_b$. Note that the amplitudes `A' of Eq.~\ref{vzdrive} and Eq.~\ref{vthdrive} can be different, as described in the following.

\section{Results}  \label{sec:results}

The simulations are run for $t_f \approx 850\ \mathrm{s}$. The integrated sonic (Alfv\'en) vertical transit time through the domain is $t_s \approx 340\ \mathrm{s}$ ($t_A \approx 400\ \mathrm{s}$), at the center of the flux tube. Thus the simulated duration is sufficient for perturbations induced by the driver to enter the corona. In the following, the results with the various drivers discussed in Section~\ref{sec:model} will be presented separately. 

\subsection{Upflows}

\begin{figure}[h]
    \centering     
        \includegraphics[width=1.0\textwidth]{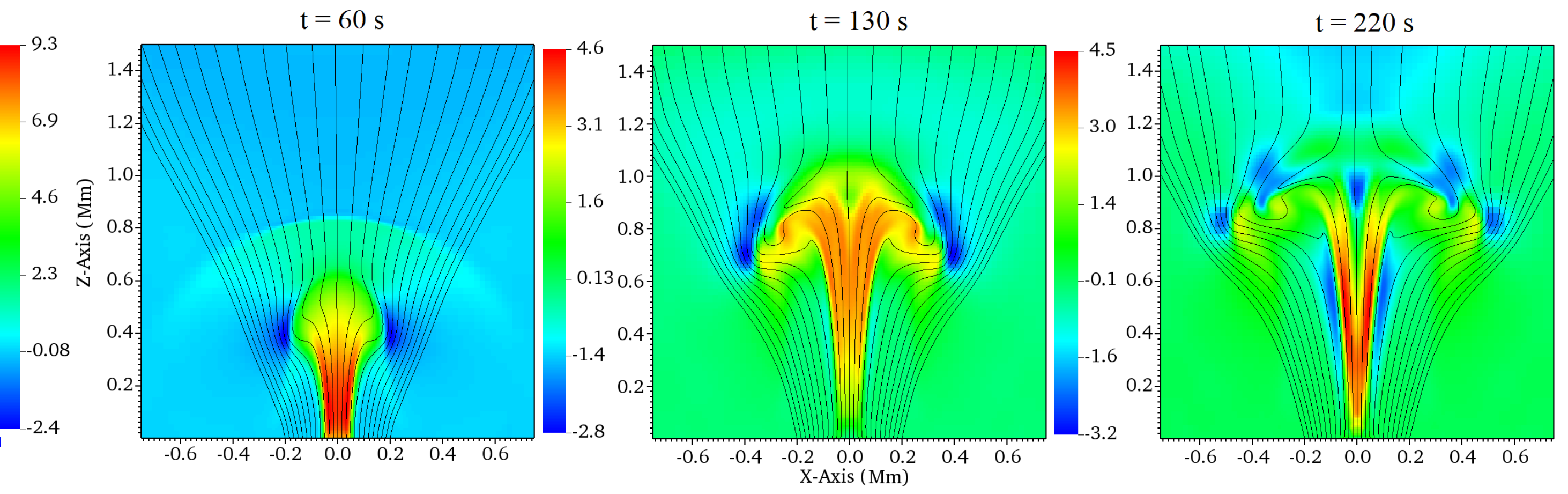}  
        \caption{Snapshots showing the vertical component of the velocity perturbation ($v_z$), in a slice along the y-axis (through $y=0$), at three different times, indicated at the top of each snapshot. Overplotted are magnetic field lines, originating from the bottom boundary. Colorbar in units of $km/s$.}
        \label{upflow_2D}
\end{figure}

\begin{figure}[h]
    \centering     
        \includegraphics[width=0.5\textwidth]{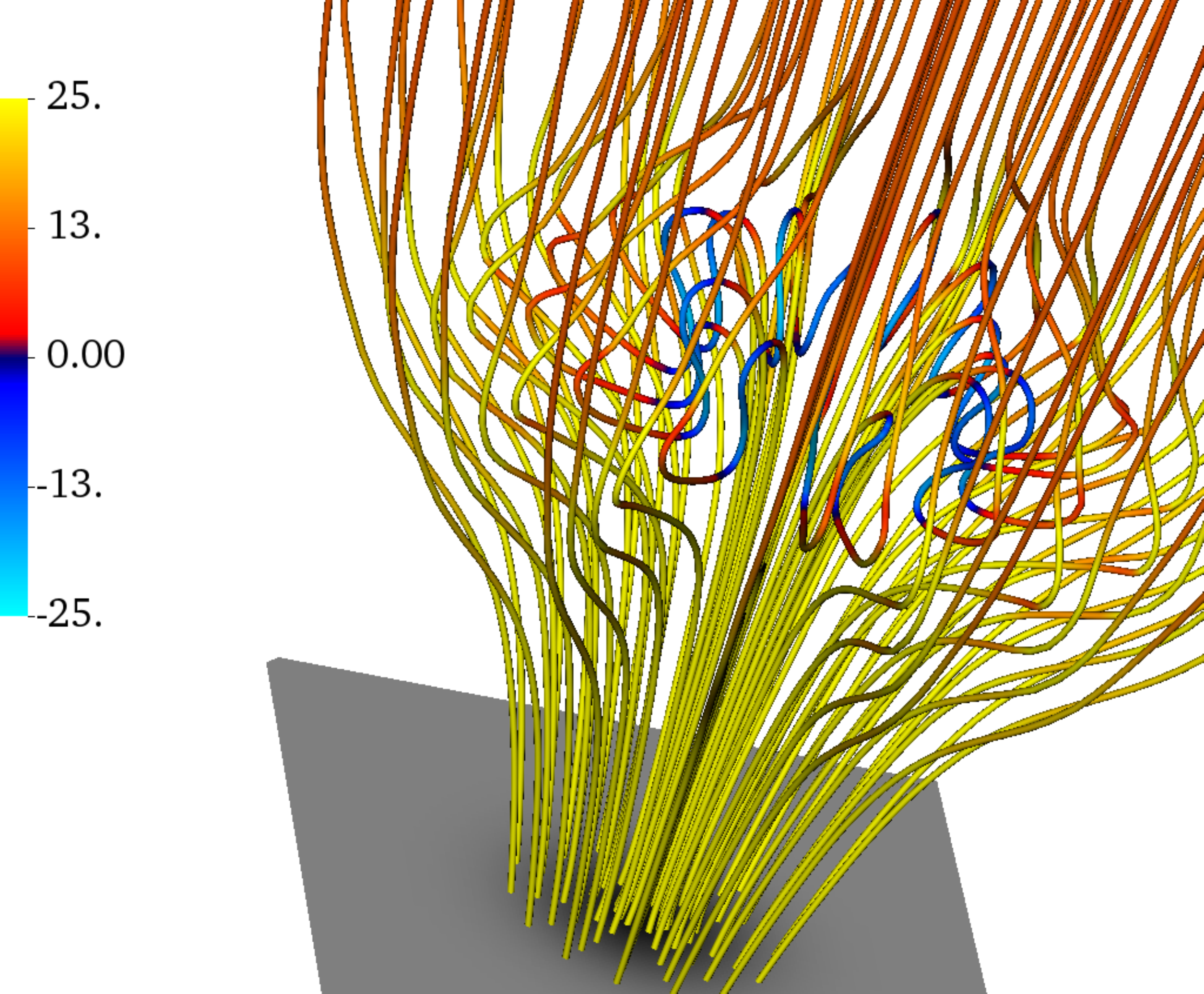}  
        \caption{3D streamplot of selected magnetic field lines at $t=220$ s (same as in Fig.~\ref{upflow_2D}), colored according to the magnitude of the vertical component of the magnetic field. Blue shades therefore represent local sunward polarity of the magnetic field. The gray panel at the bottom represents the bottom boundary. Colorbar in units of $G$.}
        \label{upflow_3D}
\end{figure}

In this setup, we employ the driver in Eq.~\ref{vzdrive}. The characteristic time of the upflow pulse is set to $\tau = 90$ s. The width of the perturbation is constrained to the extent of the intense flux tube, setting $r_b = 75\ \mathrm{km}$. The perturbation amplitude is set to $A = 10\ \mathrm{km/s}$. This corresponds on average to about $1\ M_s$ and $2\ M_A$, the sonic and Alfv\'enic Mach numbers, respectively. Although this speed is too high when compared to observed photospheric upflow speeds, which are up to $6\ \mathrm{km/s}$ \citep[e.g.,][]{2019SoPh..294...18M}, the driver is used only to induce a strong jet in the higher layers. Most of the interesting dynamics occur well above the bottom boundary, at around $1\ \mathrm{Mm}$. At these heights, the upflow speed is still about $10\ \mathrm{km/s}$, which is small when compared to, e.g. observed chromospheric jets speeds, such as in spicules \citep[e.g.,][]{2017ApJ...849L...7D}.
In Fig.~\ref{upflow_2D}, the ensuing dynamics are shown, presenting the early stages of the formation of magnetic field deflections. 
In Fig.~\ref{upflow_3D}, the three-dimensional appearance of the folded magnetic fields are shown. 
Similar structures were observed before in related numerical simulations, as in \citet{2013MNRAS.436.1268M}. 
There are a number of observations that can be made by investigating these snapshots. First of all, the folds in the magnetic field seem to be induced around the edges of the impulsively driven plasma upflow. This process is reminiscent of a Rayleigh-Taylor type instability. More precisely, as it is impulsively induced, and it is trailing behind an impulsively accelerated fast shock wave (visible at $t=60$ s in Fig.~\ref{upflow_2D}) it is possibly of Richtmyer-Meshkov type \citep[see, e.g.,][]{doi:10.1146/annurev.fluid.34.090101.162238}. The apparent switchbacks first form around the height of 500 km. These structures are advected upward by the initial pulse. However, as the pulse reaches its peak height, at around $t = 170$ s, the switchbacks cease to propagate outward. Subsequently, the switchbacks continue to evolve at the same height and ultimately unfold. \par 
We have repeated the above simulation considering different perturbation amplitudes. For higher pulse speeds, the peak height of the perturbation is increasing, and even elongated denser structures form, which protrude into the corona, resembling spicules. In all cases studied, while perturbations and waves reach into the corona, the switchbacks appear to be frozen into the denser chromospheric plasma, and do not enter the corona.    

\subsection{Vortical flows}

\begin{figure}[h]
    \centering     
        \includegraphics[width=1.0\textwidth]{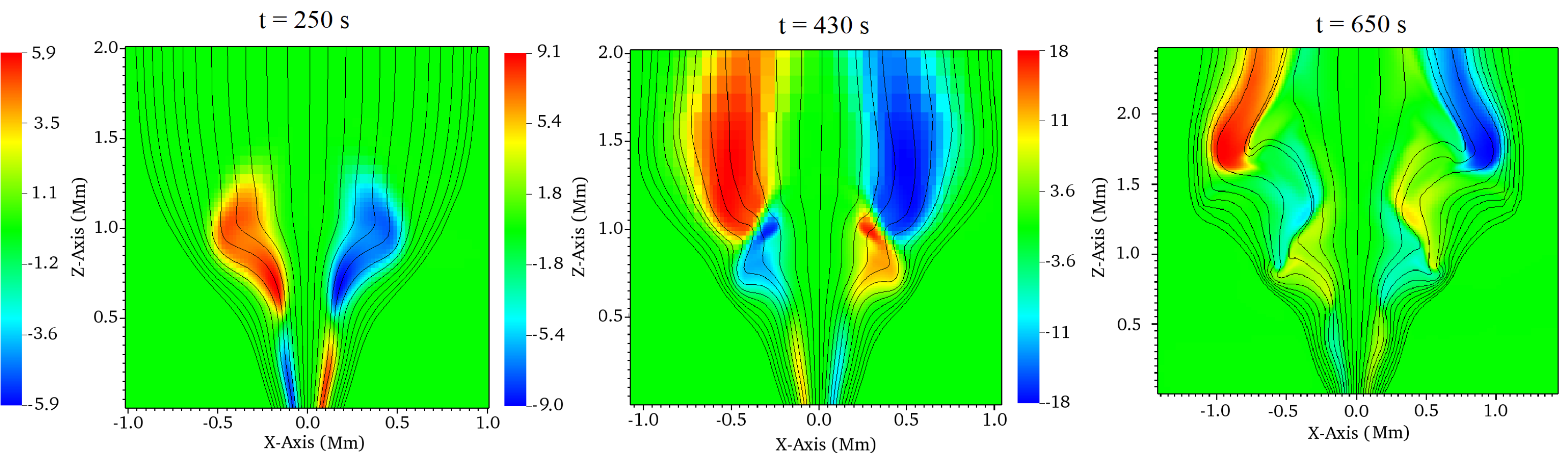}  
        \caption{Snapshots showing the LOS component of the velocity perturbation ($v_y$), in a slice along the y-axis (through $y=0$), at three different times, indicated at the top of each snapshot. Overplotted are magnetic field lines, originating from the bottom boundary. Colorbar in units of $km/s$.}
        \label{vortex_2D}
\end{figure}

\begin{figure}[h]
    \centering     
        \includegraphics[width=1.0\textwidth]{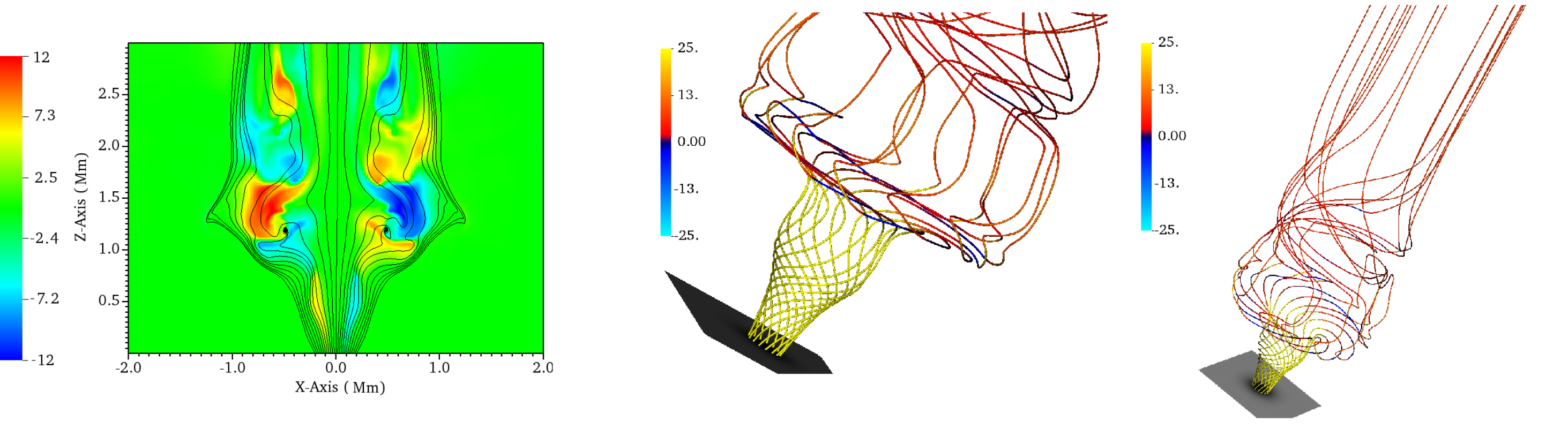}  
        \caption{Late-stage dynamics in the oscillating vortex simulation, at $t=t_f=850\ \mathrm{s}$. \textit{ Left}: Snapshot showing the LOS component of the velocity perturbation ($v_y$), in a slice along the y-axis (through $y=0$), with overplotted magnetic field lines originating from the bottom boundary. Colorbar in units of $km/s$. \textit{Centre and Right}: 3D stream plots of selected magnetic field lines, colored according to the magnitude of the vertical component of the magnetic field. Blue shades therefore represent local sunward polarity of the magnetic field. The gray panel at the bottom represents the bottom boundary. Color bars are in units of $G$.}
        \label{vortex_3D}
\end{figure}

We use Eq.~\ref{vthdrive} to drive vortical flows, and consider both a sinusoidal oscillating flow, and a decaying vortex, as described in Section~\ref{sec:model}. Same as before, we set the characteristic width of the perturbation to match the extent of the flux tube, $r_b = 75\ \mathrm{km}$. First, we present the results for an oscillating vortex flow. Oscillating vortex flows in the solar photosphere were both observed and shown to exist through numerical simulations \citep[e.g.,][]{2011A&A...533A.126M,2013ApJ...770...37K}. Although not well constrained by observations, we set the periodicity of the torsional flows to $\tau = 320\ \mathrm{s}$, which is a characteristic timescale of oscillatory flows in the photosphere (e.g., p-modes). We study different driver amplitudes, in a range $A = 3-6\ \mathrm{km/s}$, which are consistent with the observed velocities \citep[e.g.][]{2020ApJ...898..137S}. For these velocity amplitudes, initially the Alfv\'enic Mach number of the driven torsional waves is in the range 0.1 to 0.3. However, as the flux tube quickly expands and the density drops with height, the upward-propagating waves become  super-Alfv\'enic in amplitude, reaching up to $1.5\ \mathrm{M_A}$. In Fig.~\ref{vortex_2D}, the results for an oscillating driver are shown. 
By investigating Fig.~\ref{vortex_2D}, we can notice that initially the driven torsional waves are weakly nonlinear, but subsequent wavefronts are strongly affected by nonlinear wave steepening \citep[see, e.g.][]{1999JPlPh..62..219V,2017ApJ...840...64S}. This leads to strong bends in the magnetic field, which can be identified as switchbacks. In Fig.~\ref{vortex_3D}, we show the dynamics at the end of the simulation time. 
Note that over time the resulting dynamics are displaying more small-scale structure, reminiscent of transition to turbulence, resulting in many strong deflections of the magnetic field lines. 
The observed dynamics resemble those of previous studies with related setups \citep[e.g.,][]{2016MNRAS.459.2566M}. The stream plots show that while magnetic field deflections are strong, torsional waves alone result only in at a most $\pi/2$ deflection of the local magnetic field, and do no lead to complete reversals of polarity, as in the previous case with driven jets. We also note that while the chromospheric magnetic field is strongly bent by the action of the propagating torsional waves, the coronal part remains mostly straight with minimal deflections throughout the simulation, even though waves enter the corona. A possible explanation for this is the following. The Alfv\'en speed jumps in the transition region and subsequently in the corona, therefore even waves with strong deflections (i.e. large curvature, large parallel wave number), are elongated to form `long' waves (i.e. small curvature, small parallel wave number) with slight deflections. A simple analogy for a better understanding of this process is: Consider a cylindrical thread bundle (spool) on which thread is coiled or spooled up at a slow speed relative to the axis of the bundle, in a strongly helical manner. At the moment the end of the bundle is reached, consider now a fast dragging of the thread relative to the bundle axis, as to unwind the thread from the coil. The result of this fast dragging of the thread is analogous to the fast advection of the torsional waves as they enter the corona, and results practically in the straightening of the thread, or the magnetic field, respectively. \par 
Simulations with a non-oscillating, decaying vortex flow lead to similar dynamics for the same amplitudes, however the characteristic decay time is crucial to the formation of strongly bent wave fronts. Parameter studies show that if the vortical motion stops too early ($\approx\ 3$ minutes) these do not form. There is observational evidence that vortex flows in the solar photosphere could persist for a longer duration \citep[e.g.,][]{2008ApJ...687L.131B,2010ApJ...723L.139B,2011MNRAS.416..148V,2018A&A...618A..51T,2019A&A...623A.160T}.

\section{Discussion and Conclusions} \label{sec:concl}

We have investigated whether switchbacks can form as a result of different types of motions at the photospheric base of a strong magnetic flux concentration in the form of a vertical magnetic flux tube, representing e.g. a magnetic bright point. Recent solar wind measurements with the Parker Solar Probe revealed that switchbacks are not a well-defined population of magnetic field deflections, but their distribution is rather power-law-like with respect to the amount of deflection. In this sense, switchbacks, unlike the name suggests, are not necessarily full reversals of the magnetic field. The numerical model consisted of a self-similar expanding magnetic field solution, with an analytically-derived pressure-balanced atmosphere based on the \citet{2007ApJ...667.1243F} lower solar atmospheric model. The perturbations to this equilibrium were induced at the bottom boundary, consisting of either field-aligned, vertical flows, i.e., jets with upflow speeds on the order of the local sonic Mach number, or vortex flows transverse to the magnetic field resulting in torsional wave propagation. In both cases, the answer to our initial question is positive, as the ensuing non-linear processes result in strongly-deflected magnetic fields, resembling switchbacks. The details of their formation mechanisms are different for the two separate drivers, though. In the case of a jet-like driver, the switchbacks form as a result of a Rayleigh-Taylor-type instability, and roll-ups form on the sides of the protruding jet. These roll-ups can lead to true full-reversals of the local magnetic field, and to the formation of plasmoids (magnetic `bubbles') through reconnection. In the case of a vortex driver, nonlinear wave steepening of the torsional waves is responsible for the generation of strongly deflected magnetic fields. However, this mechanism cannot lead by itself to full magnetic field reversals, but is rather limited to local deflections around $\pi/2$. We note that the nature of the generated switchbacks is also different for the two drivers. The vortex driver induces torsional Alfv\'en waves which steepen nonlinearly, thus the resulting switchbacks are of Alfv\'enic nature. On the other hand, the roll-ups induced by the Rayleigh-Taylor-type instability of the driven jet are more adequately categorized as non-propagating vortex structures. \par 
While it seems likely that switchbacks are frequently generated in the lower solar atmosphere, as demonstrated in the present simulations, the other important question is whether these structures are able to leave the solar chromosphere and result in propagating or advected switchbacks into the corona. The simulations presented here suggest that switchbacks generated in the lower solar atmosphere are not likely to enter the corona. While the total simulation time is a few times longer than the wave transit times in the vertical direction, and waves are seen propagating through the corona, highly deflected magnetic fields are only present in the chromospheric layers. We offer different explanations for this observation depending on the origin of the highly deflected magnetic field structures. In the case of roll-ups or plasmoids resulting from the instability of the jet, these are frozen-in into the dense chromospheric plasma, therefore their motion is restricted to the bulk motion of the plasma. High-velocity jets of chromospheric plasma are seen protruding into the corona, and while they may contain highly deflected magnetic fields, these jets eventually fall back under the effect of gravity. In the case of highly steepened torsional waves, induced by the vortex flow driver, the strong magnetic field bends abruptly unwind as they propagate into the corona, in analogy to the unwinding of a thread spool at the fast pulling of the thread along the axis of the spool. \par 
We must stress that these conclusions are relevant under the physical model considered, that is, MHD dynamics in the absence of non-ideal non-partial ionisation effects, except of slight numerical dissipative coefficients. Thermal conduction and optically thin radiative cooling are neglected, however these non-ideal effects are mostly relevant for coronal dynamics, while in our simulations the relevant dynamics are restricted to chromospheric heights. Some of the neglected chromospheric physics, especially resistive terms leading to the violation of the frozen-in condition, such as ambipolar diffusion, might be relevant in this context \citep{1999ApJ...511..193V,2020ApJ...899L...4Y}. We leave the investigation of these effects on the dynamics of switchbacks to a follow-up study.  \par 
Furthermore, it is still an open question whether switchbacks that are already in the solar corona, either formed there by some process or entering from the lower solar atmospheric layers by some process not described by our MHD simulations, are able to propagate out to distances where these can be measured by e.g., PSP. We will investigate this possibility in the next paper of this series, by following the propagation of switchbacks embedded in realistic solar atmospheres. 
\acknowledgments

N.M. was supported by a Newton International Fellowship of the Royal Society.
D.U. is thankful for the support received through FWF project P27800. This research has received financial support from the European Union’s Horizon 2020 research and innovation program under grant agreement No. 824135 (SOLARNET) enabling D.U. a visit to Sheffield University.
R.E. is grateful to Science and Technology Facilities Council (STFC, grant number ST/M000826/1) UK and the Royal
Society for enabling this research. R.E. also acknowledges the support received by the CAS Presidents International
Fellowship Initiative Grant No. 2019VMA052 and the warm hospitality received at USTC of CAS, Hefei, where part of his contribution was made.
V.M.N. acknowledges the STFC consolidated grant ST/T000252/1 , and the BK21 plus program through the National Research Foundation funded by the Ministry of Education of Korea.

\bibliography{Biblio}{}
\bibliographystyle{aasjournal}

\end{document}